\documentclass[useAMS,usenatbib,usegraphicx]{mn2e}

\usepackage{times}

\newcommand{\smallcolsize}{0.42}
\newcommand{\singlecolsize}{0.48}

\newcommand{\sqdeg}{{\rm deg}^2}

\newcommand{\hunits}{{\rm\,km\,s^{-1}\,Mpc^{-1}}}

\newcommand{\rmodel}{r_{\rm model}}

   \newcommand{\aap}{A\&A}
\newcommand{\aj}{AJ}         \newcommand{\apj}{ApJ}
\newcommand{\apjl}{ApJ}      \newcommand{\apjs}{ApJS}
\newcommand{\mnras}{MNRAS}   
     \newcommand{\pasp}{PASP}
\newcommand{\procspie}{Proc.\ SPIE}  \newcommand{\aaps}{A\&AS}

\newcommand{\micr}{$\mu$m}

\newcommand{\newtext}[1]{{#1}}  

\title[GAMA: G02 field and DR3]
{Galaxy And Mass Assembly (GAMA): the G02 field, Herschel-ATLAS target
  selection and Data Release 3}

\author[I.K.~Baldry et al.]
{{\parbox{\textwidth}{\raggedright I.K.~Baldry,$^{1}$
J.~Liske,$^{2}$
M.J.I.~Brown,$^{3}$
A.S.G.~Robotham,$^{4}$
S.P.~Driver,$^{4,5}$
L.~Dunne,$^{6,7}$
M.~Alpaslan,$^{8,9}$
S.~Brough,$^{10}$
M.E.~Cluver,$^{11}$
E.~Eardley,$^{6}$
D.J.~Farrow,$^{12}$
C.~Heymans,$^{6}$
H.~Hildebrandt,$^{13}$
A.M.~Hopkins,$^{14}$
L.S.~Kelvin,$^{1}$
J.~Loveday,$^{15}$
A.J.~Moffett,$^{16}$
P.~Norberg,$^{17}$
M.S.~Owers,$^{18}$
E.N.~Taylor,$^{19}$
A.H.~Wright,$^{13}$
S.P.~Bamford,$^{20}$
J.~Bland-Hawthorn,$^{21}$
N.~Bourne,$^{6}$
M.N.~Bremer,$^{22}$
M.~Colless,$^{23}$
C.J.~Conselice,$^{20}$
S.M.~Croom,$^{21}$
L.J.M.~Davies,$^{4}$
C.~Foster,$^{14,21}$
M.W.~Grootes,$^{24}$
B.W.~Holwerda,$^{25}$
D.H.~Jones,$^{3}$
P.R.~Kafle,$^{4}$
K.~Kuijken,$^{26}$
M.A.~Lara-Lopez,$^{27}$
\'{A}.R.~L\'{o}pez-S\'{a}nchez,$^{14,18}$
M.J.~Meyer,$^{4}$
S.~Phillipps,$^{22}$
W.J.~Sutherland,$^{28}$
E.~van~Kampen,$^{29}$
S.M.~Wilkins$^{15}$
}}\\
\vspace{0.4cm}\\
{\parbox{\textwidth}{\raggedright 
$^{1}$Astrophysics Research Institute, Liverpool John Moores University, IC2, Liverpool Science Park, 146 Brownlow Hill, Liverpool, L3 5RF, UK
\\
$^{2}$Hamburger Sternwarte, Universit{\"a}t Hamburg, Gojenbergsweg 112, 21029 Hamburg, Germany
\\
$^{3}$School of Physics and Astronomy, Monash University, Clayton, Victoria 3800, Australia
\\
$^{4}$International Centre for Radio Astronomy Research (ICRAR), University of Western Australia, 35 Stirling Highway, Crawley, WA6009, Australia
\\
$^{5}$School of Physics and Astronomy, University of St Andrews, North Haugh, St Andrews, KY16 9SS, UK
\\
$^{6}$Institute for Astronomy, University of Edinburgh, Royal Observatory, Blackford Hill, Edinburgh EH9 3HJ, UK
\\
$^{7}$School of Physics and Astronomy, Cardiff University, Queens Buildings, The Parade, Cardiff, CF24 3AA, UK
\\
$^{8}$NASA Ames Research Center, N244, Moffett Field, Mountain View, CA 94035, USA
\\
$^{9}$NYU Center for Cosmology and Particle Physics, New York, NY 10002, USA
\\
$^{10}$School of Physics, University of New South Wales, NSW 2052, Australia
\\
$^{11}$Department of Physics and Astronomy, University of the Western Cape, Robert Sobukwe Road, Bellville, 7535, South Africa
\\
$^{12}$Max-Planck-Institut f\"{u}r Extraterrestrische Physik, Postfach 1312 Giessenbachstrasse, D-85741 Garching, Germany
\\
$^{13}$Argelander-Institut f{\"u}r Astronomie, Universit{\"a}t Bonn, Auf dem H{\"u}gel 71, 53121 Bonn, Germany
\\
$^{14}$Australian Astronomical Observatory, PO Box 915, North Ryde, NSW 1670, Australia
\\
$^{15}$Astronomy Centre, University of Sussex, Falmer, Brighton BN1 9QH, UK
\\
$^{16}$Vanderbilt University, 2401 Vanderbilt Place, Nashville, TN 37240, USA
\\
$^{17}$Institute for Computational Cosmology, Department of Physics, Durham University, South Road, Durham DH1 3LE, UK
\\
$^{18}$Department of Physics and Astronomy, Macquarie University, NSW 2109, Australia
\\
$^{19}$Centre for Astrophysics and Supercomputing, Swinburne University of Technology, P.O. Box 218, Hawthorn, VIC 3122, Australia
\\
$^{20}$Centre for Astronomy and Particle Theory, University of Nottingham, University Park, Nottingham NG7 2RD, UK
\\
$^{21}$Sydney Institute for Astronomy, School of Physics, University of Sydney, NSW 2006, Australia
\\
$^{22}$Astrophysics Group, HH Wills Physics Laboratory, University of Bristol, Tyndall Avenue, Bristol BS8 1TL, UK
\\
$^{23}$Research School of Astronomy and Astrophysics, The Australian National University, Cotter Road, Weston Creek, ACT 2611, Australia
\\
$^{24}$ESA/ESTEC, 2201 AZ Noordwijk, The Netherlands
\\
$^{25}$Department of Physics and Astronomy, 102 Natural Science Building, University of Louisville, Louisville KY 40292, USA
\\
$^{26}$Leiden University, P.O.~Box 9500, 2300 RA Leiden, The Netherlands
\\
$^{27}$Dark Cosmology Centre, Niels Bohr Institute, University of Copenhagen, Juliane Maries Vej 30, DK-2100 Copenhagen, Denmark
\\
$^{28}$Astronomy Unit, Queen Mary University London, Mile End Rd, London E1 4NS, UK
\\
$^{29}$European Southern Observatory, Karl-Schwarzschild-Str.~2, 85748 Garching, Germany
\\
}}}

\begin{document}

\date{2017 November, accepted by MNRAS}

\pagerange{\pageref{firstpage}--\pageref{lastpage}} \pubyear{2017}

\maketitle

\label{firstpage}

\clearpage
\begin{abstract}
  We describe data release 3 (DR3) of the Galaxy And Mass Assembly
  (GAMA) survey. The GAMA survey is a spectroscopic redshift and 
  multi-wavelength photometric survey in three equatorial regions 
  each of 60.0\,deg$^2$ (G09, G12, G15), 
  and two southern regions of \newtext{55.7\,deg$^2$ (G02) and 50.6\,deg$^2$ (G23)}. 
  DR3 consists of: the first release of data covering the G02 region and of 
  data on H-ATLAS sources in the equatorial regions; 
  and updates to data on sources released in DR2. 
  DR3 includes 154\,809 sources with secure redshifts across four regions. 
  A subset of the G02 region is 95.5\% redshift complete to 
  $r<19.8\,\mathrm{mag}$ over an 
  area of $19.5\,\sqdeg$, with 20\,086 galaxy redshifts, 
%% nq > 3 and survey_class > 3 and z > 0.002 and dec > -6.0 (G02TilingCat)
  that overlaps substantially with the XXL survey (X-ray) and 
  VIPERS (redshift survey). 
  In the equatorial regions, the main survey has even higher completeness
  ($98.5$\%), and spectra for about 75\% of H-ATLAS filler targets 
  were also obtained. 
  This filler sample extends spectroscopic redshifts, 
  for probable optical counterparts to H-ATLAS sub-mm sources, 
  to 0.8\,mag deeper ($r<20.6\,\mathrm{mag}$) than the GAMA main survey. 
  There are 25\,814 galaxy redshifts for H-ATLAS sources
  from the GAMA main or filler surveys. 
%% nq >= 3 and survey_class >= 1 and z > 0.002 (H-ATLAS matched to TilingCat)
  GAMA DR3 is available at the survey website (www.gama-survey.org/dr3/). 
\end{abstract}


\section{Introduction}
\label{sec:intro}

Modern day surveys designed to study galaxy evolution typically combine 
data from many wavelength regimes. Often this starts out with 
an optical imaging or spectroscopic survey, which can be a wide field
or a deep-small field, and other instruments follow suit 
adding to the available data that can be combined. 
This is useful because different phenomena dominate at different
wavelengths: young stars in the UV, older stars in the near-IR, 
dust emission in the far-IR, AGN-driven jets in the radio, 
and hot gas around AGN or in clusters of galaxies in the X-ray bands. 
Investigating the connections between these and other phenomena is
enabled by a multi-wavelength approach 
(e.g.\ \citealt{JD99,DG03,scoville07,driver16}).\footnote{List of 
abbreviations used in paper:
AAT, Anglo-Australian Telescope; 
AGN, active galactic nucleus/nuclei; 
CFHTLenS, Canada-France-Hawaii Telescope Lensing Survey; 
CFHTLS, Canada-France-Hawaii Telescope Legacy Survey; 
GALEX, Galaxy Evolution Explorer (telescope);
GAMA, Galaxy And Mass Assembly (survey); 
H-ATLAS, Herschel -- Astrophysical Terahertz Large Area Survey;
HerMES, Herschel Multi-tiered Extragalactic Survey; 
IR, infrared; 
KiDS, Kilo Degree Survey; 
PRIMUS, PRIsm MUlti-object Survey; 
SDSS, Sloan Digital Sky Survey; 
2dF, Two-Degree Field (instrument); 
2dFGRS, 2dF Galaxy Redshift Survey; 
UKIDSS, UKIRT Infrared Deep Sky Survey; 
UV, ultraviolet; 
VIDEO, VISTA Deep Extragalactic Observations (survey); 
VIKING, VISTA Kilo-Degree Infrared Galaxy (survey); 
VIPERS, VIMOS Public Extragalactic Redshift Survey; 
VISTA, Visible and Infrared Survey Telescope for Astronomy;
VST, VLT Survey Telescope; 
VVDS, VIMOS VLT Deep Survey;
WISE, Wide-field Infrared Survey Explorer (telescope); 
XMM, X-ray Multi-Mirror (telescope); 
\newtext{XXL, XMM eXtra Large (survey)}.
}
%%ASKAP, Australian Square Kilometre Array Pathfinder (radio telescope);
%%VIRCam, VISTA IR Camera;

With the advent of wide-field imagers at the European Southern
Observatory, OmegaCAM on the VST \citep{kuijken02} and the VISTA 
Infrared Camera \citep{dalton06}, large public surveys were sought. One of
these, KiDS using the VST, was approved to cover $1500\,\sqdeg$ \citep{dejong13}.  
The chosen sky areas covered the 2dFGRS \citep{colless01} in the south 
and the 2dFGRS and SDSS \citep{stoughton02} 
near the celestial equator for their spectroscopic redshifts. 
The 2dFGRS areas were originally chosen for low Galactic extinction,
i.e.\ in the Galactic caps, and for all year access from the AAT. 
The VIKING survey \citep{edge13} was designed to cover the same area of sky 
as KiDS. VIKING observations are now complete, over a final area of $1350\,\sqdeg$, 
and KiDS will cover the same area, i.e.\ 90\% of the original aim. 

In 2007, the GAMA survey was selected as a large-programme galaxy
redshift survey on the AAT, using an update to the 2dF \newtext{spectrograph} called AAOmega
\citep{sharp06}.  The motivations included an aim for high redshift
completeness to $r<19.8\,\mathrm{mag}$ for reliably selecting groups
of galaxies to measure the halo mass function, and for a general
purpose study of galaxy evolution using multi-wavelength data
\citep{driver09}. The areas chosen were primarily within the KiDS
footprint with GAMA regions now known as G09, G12 and G15 straddling
the equator, and starting later, G23 in the south 
(see Table~\ref{tab:gamaregions} for details of the GAMA regions). These four regions
were also chosen to be observed with the Herschel Space Observatory,
in the far-IR, as part of the H-ATLAS \citep{eales10}.

\begin{table*}
%\begin{minipage}{16cm}
\caption{Overview of the GAMA survey regions. 
  The southern G02 and G23 regions were not part of GAMA~I. 
  The last column provides the magnitude limits for DR3, which was selected from GAMA~II.
  The qualifier `GAMA~II' refers to the fact that a revised input catalogue was used for the 
  second phase of the GAMA survey. 
  See \citet{baldry10} for a detailed description of the GAMA~I input catalogue and
  \citet{liske15} for a description of the changes to the input catalogue for GAMA~II.
  Thus, I and II refer to two phases of target selection, and not the data releases, DR1, DR2, and DR3.}
\label{tab:gamaregions}
\centerline{\begin{tabular}{lr@{\hspace{2mm}--\hspace{2mm}}lr@{\hspace{2mm}--\hspace{2mm}}lr@{\hspace{2mm}--\hspace{2mm}}l@{\hspace{7mm}}c@{\hspace{7mm}}ccc}
\hline
Survey region & \multicolumn{2}{c}{RA range (J2000)} & \multicolumn{4}{c}{Declination\ range (J2000)} & Area & \multicolumn{3}{c}{main survey limits ($r$ band except in G23)}\\
& \multicolumn{2}{c}{(deg)} & \multicolumn{4}{c}{(deg)} & (deg$^2$) & \multicolumn{3}{c}{(mag)}\\
& \multicolumn{2}{c}{} & \multicolumn{2}{c}{GAMA~I} & \multicolumn{2}{c}{GAMA~II} & GAMA~II & GAMA~I & GAMA~II & DR3\\
\hline 
G02 & $30.2$  & $38.8$  & \multicolumn{2}{c}{--} & $-10.25$ & $-3.72$ & $55.71$ & --     & $19.8$ & $19.8$\\
G09 & $129.0$ & $141.0$ & $-1.0$  & $+3.0$       & $-2.0$   & $+3.0$  & $59.98$ & $19.4$ & $19.8$ & $19.0$\\
G12 & $174.0$ & $186.0$ & $-2.0$  & $+2.0$       & $-3.0$   & $+2.0$  & $59.98$ & $19.8$ & $19.8$ & $19.0$\\
G15 & $211.5$ & $223.5$ & $-2.0$  & $+2.0$       & $-2.0$   & $+3.0$  & $59.98$ & $19.4$ & $19.8$ & $19.8$\\
G23 & $339.0$ & $351.0$ & \multicolumn{2}{c}{--} & $-35.0$  & $-30.0$ & $50.59$ & --     & $19.2$\rlap{~($i$ band)} & --    \\
\hline
\end{tabular}}
%\end{minipage}
\end{table*}

Unfortunate delays to VST meant that GAMA target selection 
was based on SDSS for the equatorial fields, and 
an additional GAMA field was sought and chosen, G02, 
to cover the CFHTLS-W1 field \citep{Gwyn12}. 
The initial aim was to make use of the
combined lensing maps from the CFHTLenS team \citep{heymans12} and
GAMA group catalogue \citep{robotham11} based on a highly-complete
redshift survey to $r<19.8\,\mathrm{mag}$. However, this GAMA region was not
observed to a high level of redshift completeness except in an area that
overlaps with the XXL survey, XXL-N field \citep{pierre16}. Thus, while the G02 region does not have
the same homogeneous data set from the $u$-band to far-IR that have
covered the other four regions (KiDS/VIKING/H-ATLAS), it has the largest
area covered by XMM-Newton observations. Other surveys such 
as VIDEO \citep{jarvis13} and HerMES \citep{oliver12} 
cover some of the XXL-N field; 
and there are observations in the $K$-band with CFHT \citep{ziparo16} 
and 3.6 and 4.5\,\micr\ with Spitzer \citep{lonsdale03,bremer12}. 

The total sky area of the five GAMA regions is $286\,\sqdeg$.  The
first and second data releases of the GAMA survey, as well as
extensive survey diagnostics, are presented in \citet{driver11} and
\citet{liske15}. The target selection and the 2dF tiling strategy are
described in \citet{baldry10} and \citet{robotham10}, with
spectroscopic reduction and redshift measurements described in
\citet{hopkins13} and \citet{baldry14}.  Curation of and photometric
measurements using the multi-wavelength imaging data, for the four
regions excluding G02, are described in \citet{driver16}.

In DR2, data products based on spectroscopic data or redshifts were
released for targets down to $r<19.0\,\mathrm{mag}$ in G09 and G12, and 
$r<19.4\,\mathrm{mag}$ in
G15. The aim of this paper is to describe the third data release of
the GAMA survey.  In addition to the DR2 targets, this includes: data
from the G02 region; data on H-ATLAS selected targets regardless of
magnitude; data on targets in G15 to $r<19.8\,\mathrm{mag}$; 
expanded areal coverage of the equatorial regions; 
and any updates to data products since DR2.  The G02 data are described in
\S\,\ref{sec:imaging} and~\ref{sec:target-select}.  The H-ATLAS target
selection is described in \S\,\ref{sec:h-atlas}, GAMA-team data
products are described in \S\,\ref{sec:dmu}, and a summary of DR3 is
provided in \S\,\ref{sec:summary}.
Optical magnitudes were corrected for Galactic dust extinction 
using the maps of \citet{SFD98}.  

\section{G02 imaging}
\label{sec:imaging}

\begin{figure}
\centerline{
\includegraphics[width=\singlecolsize\textwidth]{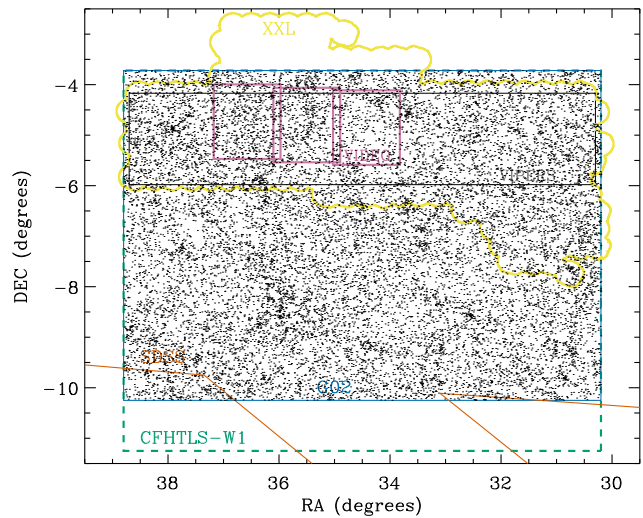}
}
\caption{Location of the GAMA/G02 field.  The blue line outlines the
  G02 region while the points represent the distribution of target
  galaxies to $r<19.8\,\mathrm{mag}$. The green dashed line outlines
  the CFHTLS-W1 field.  The yellow line outlines the XXL region. The
  orange lines show the lower boundary of SDSS coverage. The pink
  rectangles outline the VISTA-VIDEO coverage. The black rectangle 
  outlines the VIPERS coverage.}
\label{fig:location-g02}
\end{figure}

The G02 field is the region defined by $30.2<\mathrm{RA}<38.8$ and 
$-10.25<\mathrm{DEC}<-3.72$, which is a large subset, covering 87\%, 
of the CFHTLS W1 field.  As well as CFHTLS data, SDSS imaging covers 
most of the G02 field, and XXL covers about
$25\,\sqdeg$. The optical imaging surveys were used to define the
target selection, while the XXL coverage was considered when defining
the high completeness region. Figure~\ref{fig:location-g02} shows the
G02 field with the boundaries of these and other surveys.

Imaging for the CFHTLS was taken using the CFHT MegaCam instrument
\citep{boulade03},
which consists of 36 2048-by-4612 pixel CCDs. During a typical 
pointing with dithering to fill in the detector gaps,
a contiguous area of $\sim 1\,\sqdeg$ is observed. 
The W1 field was covered using 72 pointings, of which, 63 pointings
were used for the G02 region. Observations were taken
in five filters $u$, $g$, $r$, $i$, $z$ with typically 
2000\,s to 4000\,s exposures, and with seeing FWHM 
typically between $0.5''$ and $1.0''$. We used the data products 
based on the processing by the CFHTLenS team 
described in \citet{erben13}. Objects were detected 
using \textsc{SExtractor} \citep{BA96} on the $i$-band images, with
multi-wavelength photometry obtained using the
multi-image mode on PSF-matched images in all the bands
\citep{hildebrandt12}. 
A mask was also supplied by the CFHTLenS team
that removed satellite trails, optical ghosts, saturated pixels, 
diffraction spikes and other artefacts. 
An initial input catalogue of 317\,748 sources was obtained by selecting all
sources to r-band AUTO mag $<21$ that had not been masked. 

The SDSS is a set of surveys using a 2.5-m telescope
at Apache Point Observatory \citep{york00}. Imaging for SDSS used 
a large format CCD camera in drift-scan mode
with five broadband filters $u$, $g$, $r$, $i$ and 
$z$. The exposure time on source was 55\,s during
a normal drift-scan run. Gaps between the detectors
were filled in by observing with another run offset 
in the orthogonal direction to the drift scan. 
The sources and multi-wavelength photometry were 
obtained using a custom-made pipeline for SDSS
called \textsc{photo} \citep{stoughton02}. 
Here we use imaging data provided by SDSS DR8 \citep{sdssDR8}. 
An initial input catalogue of 490\,292 sources was obtained
using an SQL query to the SDSS database to 
Petrosian $r$-band mag $<21$ over a marginal superset
region (30 to 39 in RA, $-12$ to $-2$ in DEC). 
No mask was applied but a standard set of flags were 
applied that effectively removes most artefacts 
in the imaging caused by bright stars.\footnote{The selection flags for SDSS are 
given in the G02InputCat.notes file with GAMA DR3.}

\section{G02 target selection}
\label{sec:target-select}

Targets were selected from both the CFHTLenS and SDSS DR8 input catalogues,
described above, which were then merged.  SDSS
objects were matched to the nearest CFHTLenS object within a $2''$
matching radius. If an SDSS object did not have a counterpart in the
CFHTLenS catalogue then a new object was added to the merged catalogue
(e.g., this can happen for galaxies that were initially lost in the
large CFHTLenS bright star halos).  Objects could be selected for
spectroscopic targeting using photometry from either input catalogue, 
whether or not they had photometry from one or both.

For the G02 main survey, galaxies with $r < 19.8\,\mathrm{mag}$ 
after correction for Galactic dust extinction
were targeted in G02. The type of magnitudes used, 
\newtext{for this flux-limited selection}, 
were \textsc{SExtractor} AUTO \citep{Kron80,BA96} for CFHTLenS and 
Petrosian \citep{Petrosian76,stoughton02} for SDSS. 
\newtext{These both use adaptive apertures. Other magnitudes used 
were $3''$-aperture (SDSS fibre-size) magnitudes, which 
help to exclude artefacts related to the adaptive apertures, 
PSF magnitudes and profile-fitted (PSF$+$model) magnitudes. 
The differences between the latter two magnitudes for each source was used 
by SDSS as a star-galaxy profile separator.} 

Galaxies were
targeted if they met the $r < 19.8$ criterion in either the CFHTLenS
or SDSS input catalogues, with details below:
\begin{itemize}
  \item 
    For CFHTLenS, the selection criteria were objects with \textsc{SExtractor}
    $\mathrm{CLASS\_STAR} < 0.70$ and $r_{\rm auto} < 19.8\,\mathrm{mag}$.  In
    addition: targets were required to have an r-band 12 pixel 
    ($3''$, i.e., SDSS-size fibre) aperture magnitude in the range $17
    < r_{\rm fib} < 22.5$; and data in masked regions were excluded, for
    example, around bright stars.
  \item For SDSS DR8, the selection criteria were galaxies with
    $r_{\rm Petro} < 19.8\,\mathrm{mag}$. 
    Star-galaxy separation for SDSS was done
    using the method outlined by \citet{baldry10}, without the $J-K$
    measurements, using a combination of $r$-band PSF and model
    magnitudes \citep{stoughton02} as follows:
\begin{equation}
\label{eqn:sg-rband}
 r_{\rm psf} - \rmodel  >  
\begin{array}{l} 
  0.25 \\ 0.25 - \frac{1}{15}(\rmodel - 19) \\  0.15 \end{array}
\mbox{~for~} 
\begin{array}{l} 
  \rmodel < 19.0 \\ 19.0 ~~...~~ 20.5 \\ \rmodel > 20.5
\end{array} 
\end{equation}
SDSS selected targets had an SDSS $r$-band fibre
magnitude in the range $17 < r_{\rm fib} < 22.5$. A number of standard flags
were also applied to exclude artefacts.
\end{itemize}
Filler targets were selected down to $r<20.2\,\mathrm{mag}$ 
(with lower priority)
from both surveys using the same criteria, other than the change in
magnitude limit, outlined above.

Data for the targets are given in the G02TilingCat table. Targets
selected as part of the main survey can be identified using the G02
survey\_class (SC) parameter. The SC parameter takes the values: 6 for
main-survey targets selected from SDSS and CFHTLenS; 5, from SDSS
only; 4, from CFHTLenS only; and 2 for filler targets selected from
either. Visual classification was performed on a subset of $\mathrm{SC}\ge4$ 
sources, 
particularly those with discrepant photometry between the two input catalogues, to
identify artefacts, deblended parts of large galaxies and severely
affected photometry. Based on these visual checks, 290 sources were 
given an SC value of zero to indicate that they were not a target. The number of
remaining main survey ($\mathrm{SC}\ge4$) targets in G02TilingCat is 59\,285.

\begin{figure}
\centerline{
\includegraphics[width=\smallcolsize\textwidth]{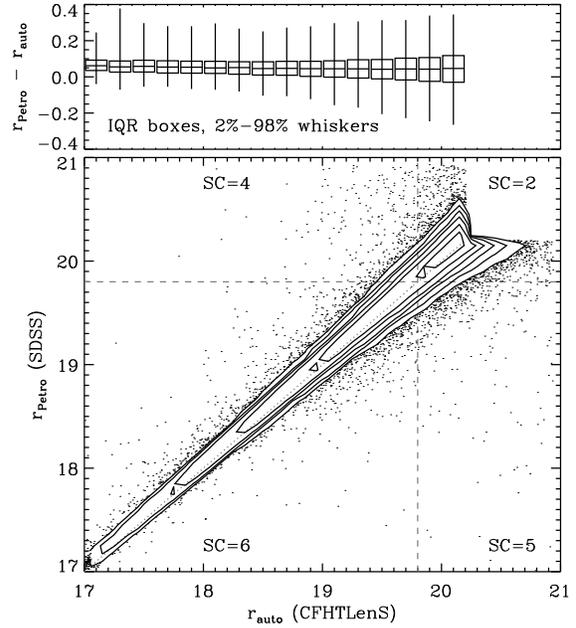}
}
\caption{Comparison between the photometry from the two G02 input catalogues 
  for sources that were identified as galaxy targets in both
  catalogues. The median value of $r_{\rm Petro} - r_{\rm auto}$ is 0.04 and
  the interquartile range (IQR) is 0.11. 
  Note there is some variation in the median offset with position on the sky; 
  this likely reflects variations in the CFHTLenS photometric calibration.  
  The dashed lines divide the different survey\_class regions.}
\label{fig:mag-compare}
\end{figure}

\begin{figure}
\centerline{
\includegraphics[width=\singlecolsize\textwidth]{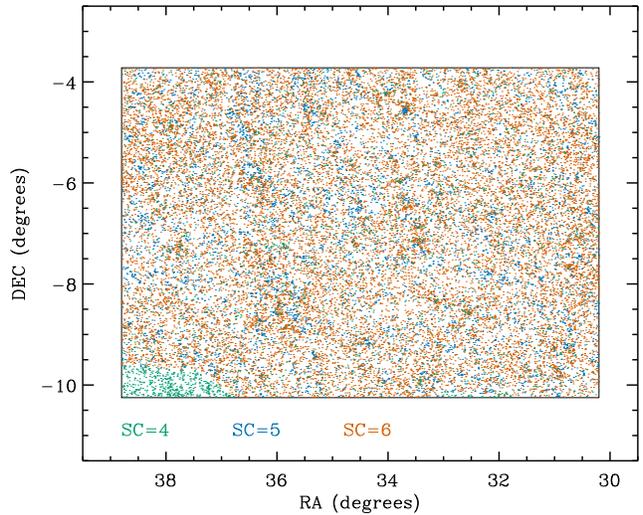}
}
\caption{Sky distribution of G02 galaxy targets colour-coded according 
  to survey\_class. The SC=4 region at the bottom left of the plot 
  corresponds to the region of G02 without SDSS coverage.}
\label{fig:survey-class-field}
\end{figure}

Figure~\ref{fig:mag-compare} shows a comparison between the selection
photometry from SDSS and CFHTLenS.  There is in general good agreement
between the two data sets, and photometric measurement codes, given
the challenging problem of galaxy photometry. Users should be aware,
however, that SC=5 and SC=4 sources dominate in certain areas of the
G02 region as shown by Fig.~\ref{fig:survey-class-field}. This is
because some areas were masked by CFHTLenS and one corner did not have
SDSS imaging.

Spectra for the targets were obtained using the AAOmega/2dF instrument on
the AAT, a multi-object fibre-fed spectrograph, excluding targets that
already had a high-quality redshift from SDSS. 
\newtext{For a single 2dF configuration (`tile'), typically 
350 fibres were allocated to targets and 
25 fibres were allocated to positions for determining a mean sky background. 
The AAOmega dual-beam setup was chosen so that spectra 
were obtained from 3750\AA\ to 8850\AA\ with a dichroic split at 5700\AA. 
The dispersion was 1\AA\ per pixel in the blue arm and 1.6\AA\ 
per pixel in the red arm. For an observation of a tile, 
data from usually three exposures and from each arm
were combined such that a single sky-subtracted 
spectrum per target was obtained \citep{hopkins13}. 
The redshifts for each spectrum were
then measured using a robust and reliability-calibrated
cross-correlation method \citep{baldry14}.}

The tiling strategy was
similar to the GAMA equatorial regions \citep{robotham10} with
priorities from high-to-low for: main-survey targets that had not been
observed spectroscopically, main-survey targets with one spectrum but
no reliable redshift, quality-control targets, and filler targets.
Clustered main-survey targets, defined as being within $40''$ of
another main-survey target, were given a boosted priority. This helps
with the strategy of obtaining high completeness, regardless of target
density, with multiple visits because of the necessity in avoiding
fibre collisions on any single visit.

\begin{figure}
\centerline{
\includegraphics[width=\singlecolsize\textwidth]{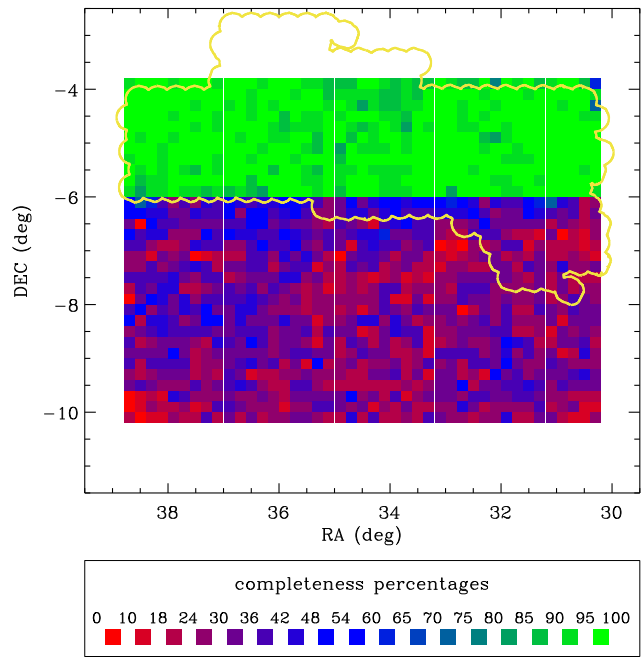}
}
\caption{Map of the redshift completeness of the G02 field.  The colour coding 
  is according to the percentage of main-survey targets 
  ($\mathrm{SC}\ge4$) with a reliable redshift ($nQ \ge 3$) 
  in areas of $0.2\degr\times0.2\degr$. 
  The yellow line outlines the XXL region.}
\label{fig:comp-map}
\end{figure}

Note that it became apparent during 2013 that the time allocated for
the GAMA survey was not going to allow completion of the G02 region to
a high completeness level. At this stage, it was decided to prioritise
the overlap with the XXL survey.  In the last season of observing,
main-survey targets north of $-6.0\degr$ were given the highest
priority though some targets between $-6.3\degr$ and $-6.0\degr$ were
observed to avoid a hard edge in completeness. As a result of this,
the redshift completeness is 95.5\% for the main survey north of
$-6.0\degr$, 46.4\% between $-6.3\degr$ and $-6.0\degr$, and 31\% south
of $-6.3\degr$, on average. 
The redshift completeness is defined as the percentage of objects in
a sample with reliable redshifts. 
Figure~\ref{fig:comp-map} shows a
completeness map of the G02 region. Note that for the area south of
$-6.0\degr$, the completeness is significantly higher for clustered
targets compared to unclustered targets. 

\begin{figure}
\centerline{
\includegraphics[width=\singlecolsize\textwidth]{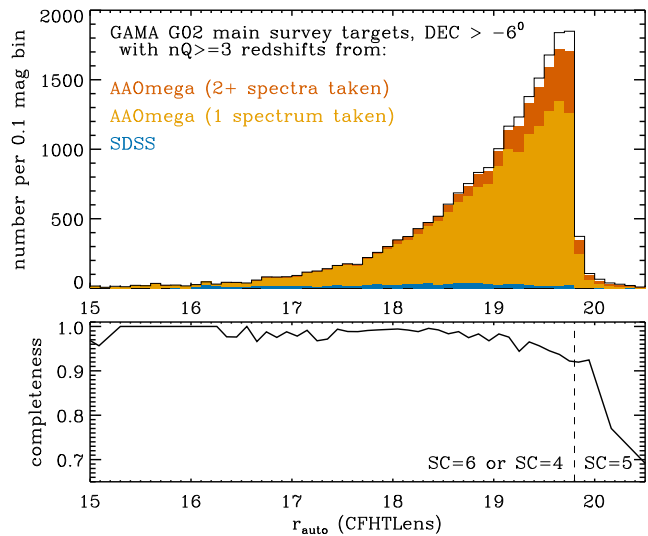}
}
\caption{Magnitude distribution and redshift completeness of the
  main-survey targets in the G02 field north of $-6.0\degr$.  The
  colour coding in the upper panel shows targets with reliable
  redshifts from SDSS and AAOmega, and whether or not more than one
  spectrum were taken for the GAMA survey. The main-survey targets 
  to the right of the dashed line were selected because of their 
  SDSS magnitudes (SC=5, see Fig.~\ref{fig:mag-compare}).}
\label{fig:completeness-mag}
\end{figure}

It is clear that the $19.5\,\sqdeg$ area north of $-6.0\degr$ has the
fidelity required for a robust group catalogue and other clustering
measurements, however, care must be taken to understand the effect of
combined SDSS and CFHTLenS selection.  There are 21\,152 main-survey
targets, of which 20\,200 have a reliable redshift with $nQ \ge 3$
(completeness of 95.5\%; $nQ$ is the redshift quality flag as 
defined in \citealt{liske15}).  The completeness is 96.2\% for main-survey
targets that have a CFHTLenS $r$-band AUTO mag measurement.
Figure~\ref{fig:completeness-mag} shows the magnitude distribution of
targets and the redshift completeness versus $r_{\rm auto}$.  The
magnitude distribution is also shown divided according to source of
redshift: SDSS or AAOmega, and in the latter case whether more than one
spectrum was taken. This demonstrates that in order to obtain high
completeness at the faint end, it is necessary to observe many of the
targets twice. This can compensate for variable conditions and fibre
throughput, as well as allowing coadding of two spectra to increase
the signal-to-noise ratio.

\begin{figure}
\centerline{
\includegraphics[width=\singlecolsize\textwidth]{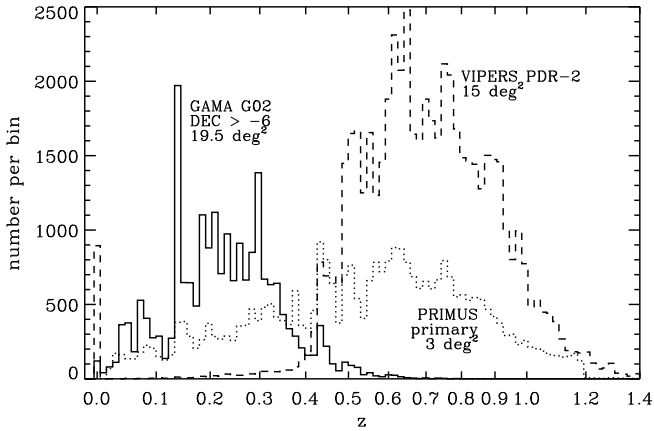}
}
\caption{Redshift histograms for the G02 area, ${\rm DEC} > -6.0\degr$,
  comparing GAMA (solid line), VIPERS (dashed line) and PRIMUS (dotted line). 
  The bin spacing is 0.01 in $\ln(1+z)$.}
\label{fig:z-histo}
\end{figure}

Figure~\ref{fig:z-histo} shows the redshift histogram of the GAMA-G02
high-completeness area. Also shown are the histograms for 
VIPERS \citep{garilli14} and PRIMUS \citep{coil11}, which covers part of G02. 
VIPERS uses photo-$z$ selection to target galaxies
at $z> 0.5$, thus, there are almost no common targets between
GAMA and VIPERS (which covers $\sim15\,\sqdeg$). In consideration of matching
with XXL, for example, this leads to the situation where the structures 
are well defined by GAMA at $z<0.35$ and by VIPERS at $z>0.5$,
with a redshift coverage gap. PRIMUS only fills this over a 
significantly smaller area.  
There are also 11\,000 redshifts available from VVDS \citep{lefevre13}, 
{\em not shown}, 
over about $0.6\,\sqdeg$ with a redshift interquartile range of 
0.54--1.12.

The G02 group catalogue was constructed using the same code as
presented in \citet{robotham11}, primarily over the highly complete
region of G02 (declination $> -6\degr$). To make use of the less
complete redshifts available below declination $-6$, we made a hard
cut at declination $> -6.3\degr$ but use $-6$ as the border flag
within the group finding code. To be consistent with the GAMA
equatorial regions, we use the SDSS $r_{\rm Petro}<19.8\,\mathrm{mag}$ selected
targets only ($\mathrm{SC}\ge5$). After redshift cuts, this
results in 20\,029 galaxies available for group finding.  The
resultant group catalogue has 2\,540 `groups' with two or more
members. We compute the same standard group statistics as per
\citet{robotham11}, with the same halo mass and group luminosity
calibrations.

It is important to note that the number of galaxies linked in each group, 
NFOF, does not have a physical meaning because it is defined using 
a magnitude-limited sample. 
Here, we compute the richness using a density-defining population (DDP)
that has $M_r < -21\,\mathrm{mag}$.\footnote{We assume 
 a flat $\Lambda$CDM cosmology with $H_0=70\hunits$ and $\Omega_{m,0}=0.3$
 for absolute magnitudes and distances.} 
The richness ($N$) is the the number of DDP galaxies in a group. 
The absolute magnitude is given by 
\begin{equation}
  M_r = r_{\rm Petro} - 5 \log (D_L / 10{\rm\,pc}) - k_r + 1.03 z  \mbox{~~~,}
 \label{eqn:mr}
\end{equation}
with k-corrections using \textsc{kcorrect} \citep{BR07} and with 
an evolution factor of 1.03 from the $Q_e$ value derived by \citet{loveday15}
(table~3 of their paper).
Given the spectroscopic limit of $r<19.8\,\mathrm{mag}$, 
the DDP galaxies are volume 
limited to $z<0.28$, though we assume the richness values are reliable 
to $z<0.3$. 

\begin{figure*}
\centerline{
\includegraphics[width=0.8\textwidth]{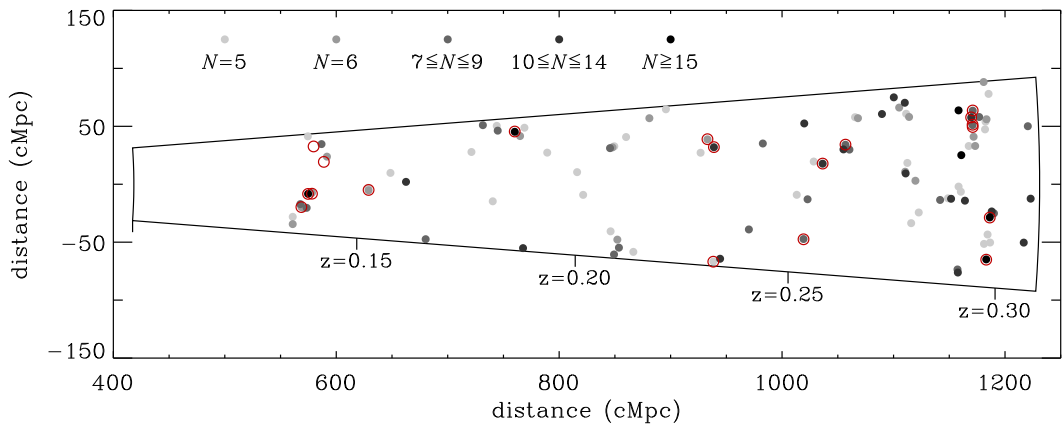}
}
\caption{Cone plot of the GAMA groups in the G02 high-completeness 
  region ($-6.0 < {\rm DEC} < -3.72$, the two dimensions
  shown correspond to RA and redshift). The GAMA groups with 
  richness $N \ge 5$ are shown with filled circles, shaded
  according to the number of galaxies with $M_r < -21\,\mathrm{mag}$. The 
  XXL clusters from the bright cluster sample of \citet{pacaud16} 
  are shown with red circles.}
\label{fig:cone-g02}
\end{figure*}

The number of `rich' groups, $N \ge 5$, in the high-completeness G02 region 
(DEC $> -6$, $19.5\,\sqdeg$), and in the redshift range $0.1 < z < 0.3$, is 98. 
This gives a number density of $3.0 \times 10^{-5}\,\mathrm{Mpc}^{-3}$. 
This is higher than the average for the equatorial regions but consistent
with cosmic variance. 
For the equatorial regions divided into 
nine $20\,\sqdeg$ areas, the number density ranges from 
$2.1 \times 10^{-5}\,\mathrm{Mpc}^{-3}$ to $3.4 \times 10^{-5}\,\mathrm{Mpc}^{-3}$.
Figure~\ref{fig:cone-g02} shows the distribution of 
the G02 rich groups in RA and redshift as a `cone plot'. 

The groups and clusters from the XXL bright 100 sample \citet{pacaud16},
with $L_{500,{\rm MT}} > 10^{43} {\rm erg\,s}^{-1}$,\footnote{$L_{500,{\rm MT}}$ is the estimate
of the X-ray luminosity within a radius within which the average mass density
of the cluster equals 500 times the critical density of the Universe.}
were matched to GAMA groups, taking the richest group within 6 arcmin and 800\,km/s 
(between the XXL-designated redshift and median redshift of GAMA galaxies in a group). 
All but two XXL groups at redshifts $z<0.4$ were matched to $N\ge5$ GAMA groups. 
The locations of XXL bright groups are also shown in Fig.~\ref{fig:cone-g02}.
The G02 region can thus be used to determine the optical properties of 
X-ray selected groups, or vice versa.

\section{H-ATLAS target selection in the equatorial regions}
\label{sec:h-atlas}

The H-ATLAS was the largest open time survey completed with the
Herschel Space Observatory \citep{eales10}.  This survey observed
$600\,\sqdeg$ of sky including four of the GAMA regions, with imaging
in five bands from 100\,\micr\ to 500\,\micr. The $4\sigma$ limit at
250\,\micr\ is about 30\,mJy \citep{valiante16}. The FWHM of the PSF is
about $18''$, which is significantly larger than the optical
images. One technique developed to identify counterparts was a
likelihood-ratio technique that assigns a reliability $R$ to nearby 
SDSS sources \citep{smith11}.

The H-ATLAS chose the GAMA equatorial regions in order to
increase the number of available redshifts matched to H-ATLAS sources
compared to, for example, SDSS or 2dFGRS.  This is because of the
higher median redshift of GAMA compared to the shallower redshift
surveys. The GAMA main survey in the equatorial regions is primarily
$r_{\rm Petro} < 19.8\,\mathrm{mag}$ but also includes about 2000 targets from
$z$-band and $K$-band selection \citep{baldry10}. In February 2011,
L.\ Dunne provided the GAMA team with a preliminary
cross-identification of H-ATLAS sources with SDSS imaging based on the
method of \citet{smith11}. Any matches (reliability $R> 0.8$), that were not in
the GAMA main survey, were included as AAOmega filler targets if they
satisfied $\rmodel < 20.6\,\mathrm{mag}$, passed the GAMA star-galaxy
separation and were not in masked regions.

The observations of the GAMA equatorial regions with AAOmega went well
such that nearly all of the main survey targets had a spectrum taken, and
this meant that 86\% of H-ATLAS filler targets had a spectrum taken as
well. This happened because, toward the end of the observations, we
could not fill every tile with main survey targets that either did not
have a reliable redshift measurement or were not already observed twice.

Subsequent to the selection of filler targets in 2011, the H-ATLAS team had
obtained some additional Herschel imaging, improved reductions, and
improved the cross-identification. The cross-identifications with SDSS
are described in \citet{bourne16}. Thus, the identifications that are
given in the H-ATLAS data release are not the same as used for the
GAMA filler targets. An assessment of the completeness issues
associated with using the H-ATLAS data release and GAMA redshifts is
described below.

We selected an H-ATLAS-GAMA sample as follows.\footnote{We used the
  data table HATLAS\_DR1\_CATALOGUE\_V1.2 from
  http://www.h-atlas.org/public-data/download} The H-ATLAS optical
identifications were matched to the GAMA input catalogue, and the
tiling catalogue that contains redshift information. Sources were
selected that: were in the GAMA equatorial regions, passed the
$r$-band star-galaxy separation (Eq.~\ref{eqn:sg-rband}), had $\rmodel
< 20.6\,\mathrm{mag}$, 
were not masked by GAMA, had H-ATLAS-SDSS cross-matching reliability $R > 0.8$,
and had $m_{250} < 12.72\,\mathrm{mag}$ 
(AB magnitudes from best flux estimate in
the 250-\micr\ band; flux density $> 29.65\,{\rm mJy}$).  
The completeness of H-ATLAS detections is significantly lower 
at fainter 250-\micr\ magnitudes \citep{valiante16}.

\begin{figure*}
\centerline{
\includegraphics[width=0.8\textwidth]{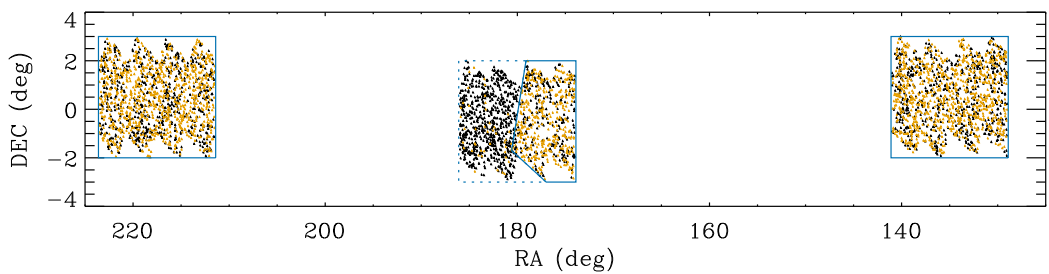}
}
\caption{H-ATLAS selected galaxies in the GAMA regions with 
  $20.0 < \rmodel < 20.6$.  The black points represent sources that were
  {\em not} GAMA targets, while the orange points represent sources
  that were GAMA targets.  The blue solid lines outline the areas
  where there was H-ATLAS coverage prior to February 2011, and 
  therefore H-ATLAS selection was available for the GAMA filler sample.
  Saw-tooth boundaries are limits of the H-ATLAS coverage arising from
  Herschel observing constraints.}
\label{fig:hatlas-coverage}
\end{figure*}

Figure~\ref{fig:hatlas-coverage} shows the distribution of these H-ATLAS
selected galaxies with $\rmodel > 20\,\mathrm{mag}$.  Also shown is whether or
not they were a GAMA target.  This demonstrates that the Herschel
imaging over half of G12 was not available for filler selection in
2011.  Only the targets that were within the blue solid outlines shown on
Fig.~\ref{fig:hatlas-coverage} were selected for this reason. 
This left an H-ATLAS-GAMA sample of 20\,380 sources.

\begin{figure}
\centerline{
\includegraphics[width=\singlecolsize\textwidth]{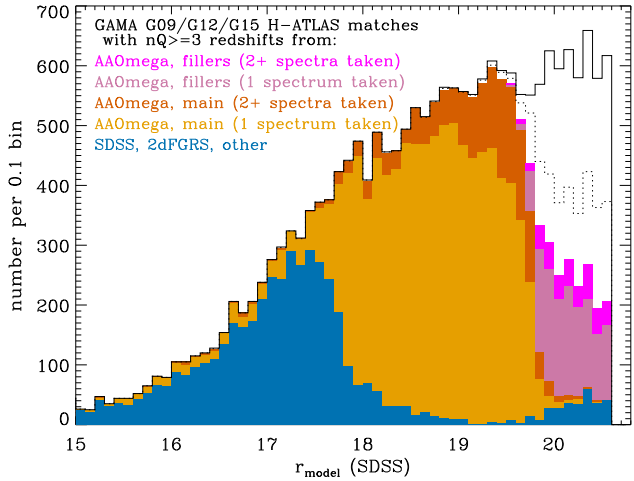}
}
\caption{Histogram showing sources of redshifts to the H-ATLAS
  matched sample as a function of magnitude. The solid line shows
the H-ATLAS-GAMA selected sample described in the text. The dotted
line shows sources that either have a reliable redshift or were
a GAMA target. The coloured filling is according to source of 
  redshift.}
\label{fig:hatlas-redshifts}
\end{figure}

\begin{figure}
\centerline{
\includegraphics[width=\smallcolsize\textwidth]{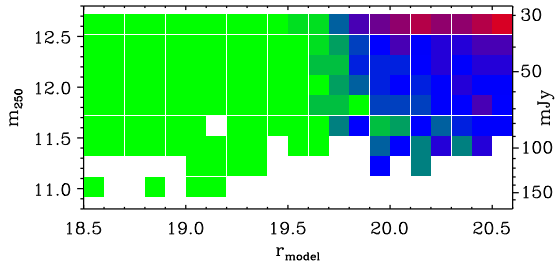}
}
\caption{Redshift completeness of the H-ATLAS-GAMA sample as a function
  of 250-\micr\ magnitude and $\rmodel$. The colour scale representing 
  completeness percentage is the same as in Fig.~\ref{fig:comp-map}.}
\label{fig:hatlas-comp-map}
\end{figure}

Figure~\ref{fig:hatlas-redshifts} shows a histogram of 
the sample in $\rmodel$, with colour filling 
where there are reliable redshifts. The redshift
completeness is high for $\rmodel < 19.6\,\mathrm{mag}$ but drops off
at fainter magnitudes (this would be a sharp drop 
at 19.8 in $r_{\rm Petro}$ because of the GAMA main
survey selection). Figure~\ref{fig:hatlas-comp-map} 
shows how the completeness depends on $m_{250}$ and
$\rmodel$. This makes it clear there is a region of low 
completeness at faint magnitudes in both filters. 
This can be understood in terms of separate effects
associated with targeting completeness and redshift 
success rate. 

\begin{figure}
\centerline{
\includegraphics[width=\smallcolsize\textwidth]{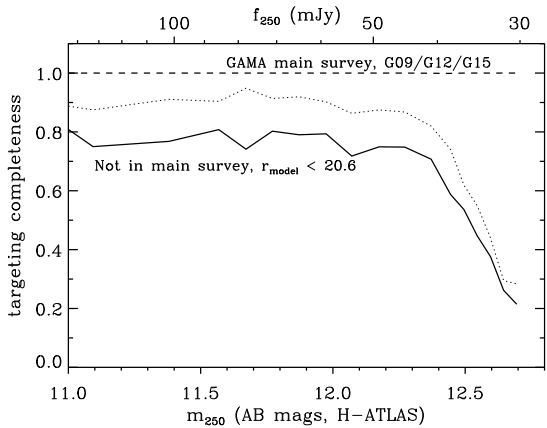}
}
\caption{Targeting completeness for the H-ATLAS-GAMA sample as a 
function of $m_{250}$. The solid and dashed line show the completeness
for non-main survey and main survey sources. The dotted line shows the
fraction of the non-main survey sources that were assigned as filler targets
in February 2011.}
\label{fig:targ-comp}
\end{figure}

Targeting completeness is defined as the percentage of objects in a sample
that have been observed spectroscopically. 
This is essentially 100\% for targets that
are in the main survey (15\,453), but depends on $m_{250}$ for non-main survey
targets (4\,927) as shown in Fig.~\ref{fig:targ-comp}.  From this, it is
clear that the targeting completeness is high for $m_{250} <
12.35\,\mathrm{mag}$ and drops off significantly at fainter magnitudes.  This
reflects the change in the cross-identifications between GAMA filler
selection and the H-ATLAS data release \citep{bourne16}.

\begin{figure}
\centerline{
\includegraphics[width=\smallcolsize\textwidth]{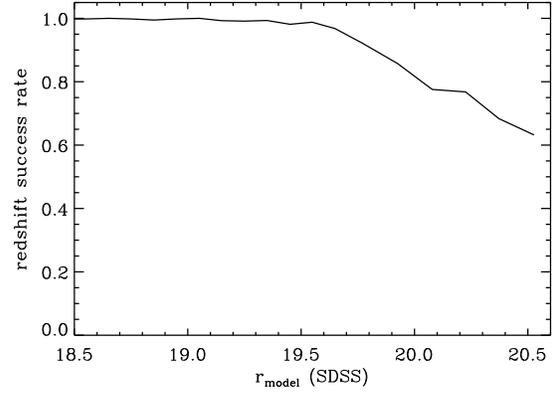}
}
\caption{Redshift success rate as a function of $\rmodel$ for the 
  H-ATLAS-GAMA sample.}
\label{fig:z-success}
\end{figure}

The other feature of the redshift completeness map shown in 
Fig.~\ref{fig:hatlas-comp-map} is the drop off toward fainter 
$\rmodel$. This is caused by a decrease in the redshift success rate, 
which is defined as the percentage of spectroscopically-observed targets 
that have a reliable redshift measurement, 
as shown in Fig.~\ref{fig:z-success}. 
To summarize, the redshift completeness map of the H-ATLAS-GAMA 
sample in Fig.~\ref{fig:hatlas-comp-map} can be explained in terms of: 
100\% targeting completeness for main survey sources; 
about 75\% targeting completeness for non-main survey sources 
at $m_{250} < 12.35\,\mathrm{mag}$ with a drop off fainter than this; 
and for all sources, a general decrease in redshift success rate with 
$\rmodel$ fainter than 19.5. 

\begin{figure}
\centerline{
\includegraphics[width=\smallcolsize\textwidth]{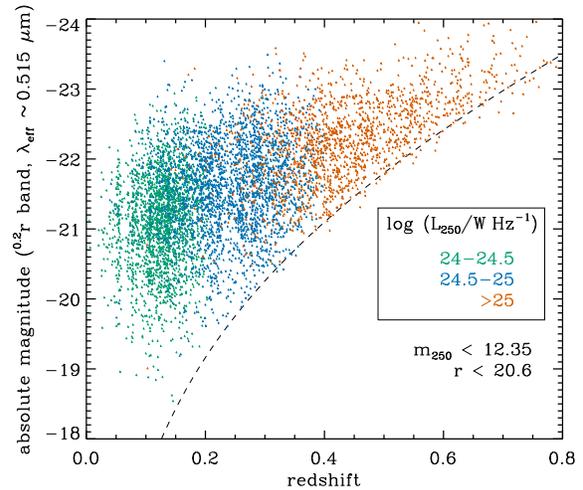}
}
\caption{Demographics of the H-ATLAS-GAMA sample to $r_{\rm
    model}<20.6\,\mathrm{mag}$ and $m_{250} < 12.35\,\mathrm{mag}$ 
  with spectroscopic redshifts.
  The dashed line shows the $r$-band magnitude limit using an average
  k-correction as a function of redshift. The points are colour coded
  according to their 250-\micr\ rest-frame luminosity, which is
  estimated using $m_{250}$ and an average k-correction as a function
  of redshift.}
\label{fig:hatlas-absmag-z}
\end{figure}

To show the demographics of the H-ATLAS-GAMA sample, we select sources
with galaxy spectroscopic redshifts less than 0.8, 
$m_{250} < 12.35\,\mathrm{mag}$,
a random selection of 75\% of the main survey, and all the fillers to
$r_{\rm model} < 20.6\,\mathrm{mag}$. This is unbiased within these magnitude limits
other than the redshift success rate variation
(Fig.~\ref{fig:z-success}).  The distribution of this sample in
visible-band absolute magnitude versus redshift is shown in
Fig.~\ref{fig:hatlas-absmag-z}, with colour coding according to far-IR
luminosity.  This clearly demonstrates the increase in the number
density of the most luminous far-IR galaxies ($L_{250}>10^{25}{\rm
  W\,Hz}^{-1}$; cf.\ \citealt{dye10,guo14}). The filler sample 
is particularly useful for selecting luminous far-IR galaxies 
at $0.4 < z < 0.8$. 

%% So you're reading the comments embedded in the LaTeX file

%% START DMU and DR3 descriptions %%

\section{Data management units}
\label{sec:dmu}

The GAMA database is organised into data management units (DMUs). 
Here they are briefly introduced with any significant updates noted. 

\subsection{Spectroscopic, redshift and input catalogue DMUs}

\subsubsection{SpecCat}

This DMU includes the spectra and redshifts obtained from the AAT, and
tables combining the redshifts from all the curated spectroscopic data
for the GAMA survey.  In DR2, the redshifts from AAT spectra were
obtained using the semi-automatic code \textsc{runz} \citep*{SCS04}
with user interaction.  For DR3, the primary choice of redshift has
been updated; these are now from the automatic code \textsc{autoz}
\citep{baldry14} except for some broadlined-AGN spectra.

A detailed description of the redshift procedure using \textsc{runz},
and other spectroscopic survey procedures, are given in \S\,2 of
\citet{liske15}, while the \textsc{autoz} method and its calibration
are described in detail in \citet{baldry14}.  An analysis of the
survey redshift completeness, and a comparison between redshift
accuracy from \textsc{autoz} and \textsc{runz}, are described in \S\,3
of \citet{liske15}.

\subsubsection{ExternalSpec}

Spectra and redshifts were curated from 10 other surveys 
within the GAMA regions. These external spectroscopic surveys are 
used for 12\% of the best redshifts of the main-survey targets
(equatorial and G02 regions). 
This DMU is described in \S\,2.7 of \citet{liske15}, and the list of surveys
is given in table~2 of that paper. 

% minor DMU that is not in DR3
%\subsubsection{LTObs}
%
%Spectra of twenty bright sources, that were not targeted by the AAT
%because of their fibre magnitudes, were obtained with the Liverpool
%Telescope. The observations and reduction are described in \S\,2.8 
%of \citet{liske15}. 

\subsubsection{EqInputCat}

This DMU includes the input catalogue and the tiling catalogue 
for GAMA equatorial regions, G09, G12 and G15. 
The input catalogue was derived from SDSS. 
Previous versions of the tiling catalogue were 
used to track redshift completeness during the observations. 
The current version is designed to be used 
as a starting point for selecting well defined samples,
from the best redshifts for each source and information 
on target selection. 
Visual classifications were made for a significant fraction
of sources in order to identify deblends and artefacts.

A detailed description of the procedure for defining
the input catalogue and selecting targets is given 
in \citet{baldry10}. Updates to the DMU are described 
in \S\,2.9 of \citet{driver11} and 
\S\,2.1 of \citet{liske15}.

\subsubsection{G02InputCat}

The input catalogue and tiling catalogue for the G02 
region are described in \S\,\ref{sec:imaging} and 
\S\,\ref{sec:target-select} of this paper. 

\subsubsection{SpecLineSFR}

This DMU contains three tables with line flux and equivalent width
measurements for GAMA~II spectra, 
which have $nQ > 1$ and $0.002 < z < 1.35$. As well as providing the
additional measurements for the G02 region, there are several key
differences between this DMU version and an earlier version used in DR2 
(\citealt{gunawardhana13}, \citealt{hopkins13}, \S\,5.1.10 of \citealt{liske15}). 
First, this DMU provides line
fluxes and equivalent widths (EWs) measured in a consistent manner for spectra from 
several surveys that used AAOmega, SDSS, 2dF or 6dF; 
although we note that flux measurements are only useful for the SDSS and AAOmega spectra
that were flux calibrated. Second, uncertainties are estimated for
the line flux and EWs by propagating the formal uncertainties on the
fitted parameters. Third, we include direct summation EW measurements
for various line species, as well as estimates of the strength of the
4000\AA\, break, $D_{4000}$. Fourth, we include more complicated
two-Gaussian fits to the $H\alpha$ and $H\beta$ emission lines where
the second component accounts for broad emission or stellar
absorption. We include various model selection techniques to determine
where more complicated models are favoured over simple ones. The
fitting, line measurements and model selection techniques are
described in detail in \citet{gordon17}.

\subsubsection{LocalFlowCorrection}

This DMU provides corrections from the 
heliocentric redshifts to the 
cosmic-microwave-background frame, and
uses the flow model of \citet{tonry00} to provide
redshifts primarily corrected for Virgo-cluster 
infall at low redshift. The procedure for,
and the effect of, this are described
in detail in \S\,2.3 of \citet{baldry12}.

\subsection{Image analysis DMUs}

\subsubsection{ApMatchedPhotom}
\label{ApMatchedPhotom}

This DMU provides AUTO (Kron) and Petrosian photometry, as well as
other \textsc{SExtractor} outputs \citep{BA96} for the $ugrizZYJHK$ bands. 
The DMU is described in detail in \citet{driver16}.  This
outlines the processing of the VISTA VIKING data \citep{Sutherland12,edge13} 
in detail, along with aperture-matched photometry from SDSS $ugriz$ and VISTA
$ZYJHK$. As part of the process all data were smoothed to a common seeing FWHM
of $2''$ to ensure accurate colour measurements. Earlier versions of
ApMatchedPhotom were based on SDSS and UKIDSS \citep{warren07}) data and 
later versions on SDSS and VISTA VIKING data where a significant
improvement was seen in the near-IR colours in particular. 
ApMatchedPhotom has been superseded by the in-house \textsc{lambdar} code \citep{wright16} 
but is included in this release for completeness and verification of earlier publications.

\subsubsection{SersicPhotometry}

The S\'ersic Photometry DMU provides single-component \citet{Sersic68}
fit results for sources across the GAMA equatorial
survey regions. Independent fits are provided for each galaxy in each
of the SDSS $ugriz$, UKIDSS $YJHK$ and VISTA $ZYJHK$
passbands. Galaxy models are constructed using \textsc{sigma} v1.0-2 (Structural
Investigation of Galaxies via Model Analysis, \citealt{kelvin12}).
\textsc{sigma} is a wrapper around several contemporary astronomy tools including
Source Extractor \citep{BA96}, PSF Extractor \citep{Bertin11}
and GALFIT~3 \citep{peng10galfit}, as well as analysis and processing
code written in the open-source R programming language. In addition
to standard GALFIT outputs, several additional value-added parameters
are also output, such as truncated S\'ersic magnitudes and central surface
brightnesses. Further details on the fitting process and outputs may
be found in \citet{kelvin12}.

\subsubsection{GalexPhotometry}

\newtext{This DMU contains catalogues for 
the NUV and FUV fluxes of each GAMA galaxy
derived using three different methods. 
These are `simple match photometry' (nearest neighbour), 
`advanced match photometry' and 
`curve-of-growth (CoG) photometry'. 
In the second case,  UV flux from the GALEX 
sources is distributed among the GAMA objects based on knowledge of the
positions and sizes of the objects involved. 
In the CoG case, surface photometry is performed on the GALEX images 
at the optically-defined location of each GAMA galaxy, 
The procedures are 
extensively described in \S\,4.2 of \citet{liske15}.}

%% Description of one of the catalogues. 
%This catalogue contains the result of an advanced matching procedure
%between the SDSS-based input catalogue (in the equatorial GAMA regions) and
%photometry from the GALEX UV space telescope. 
%The matching procedure takes into account the possibility
%of multiple matches between GALEX and GAMA objects and seeks to
%reconstruct the original UV flux of a given GAMA object.
%This catalogue only includes matched GAMA objects with S/N $\ge$ 2.5, 
%in GALEX-NUV band, in the redistributed flux. 

\subsubsection{WISEPhotometry}

This DMU provides the photometry from imaging with the 
Wide-field Infrared Survey Explorer for GAMA sources. 
The construction of the catalogue follows the methodology described in \citet{cluver14}, in
particular identifying and measuring resolved sources. Photometry of GAMA galaxies not resolved by
WISE is taken from the AllWISE Catalogue; here the ``standard aperture'' photometry (w*mag) is used,
and not the profile-fit photometry (w*mpro). This is to account for the sensitivity of WISE when
observing extended, but unresolved sources \citep{driver16}. Further details can be found in
\citet{jarrett17}. All photometry in the DMU has been corrected to reflect the updated
characterisation of the W4 filter as described in \citet*{BJC14}.

%\subsubsection{HATLASPhotometry}

%\subsubsection{XXLPhotometry}

%\subsubsection{PanchromaticPhotom}
%\label{sec:PanchromaticPhotom}

%% Description of DMU not in DR3
%The PanchromaticPhotom DMU is a convenient combination of key redshift and
%flux data from SDSS, VIKING, GALEX, WISE and Herschel-ATLAS. 
%Specific flux columns were extracted from the associated DMUs,
%extinction corrected where appropriate, and the data converted to Janskys 
%in preparation for use with \textsc{magphys} \citep{dCCE08}. 
%The process of constructing the
%PanchromaticPhotom DMU is detailed in \citet{driver16}, which includes
%comparison of adjacent colour distributions (e.g., SDSS $z$ -- VISTA $Z$) to
%literature photometry, demonstrating an improvement across survey boundaries
%as indicated by narrower and peakier colour distributions. The astrometry was
%also verified and the 1$\sigma$ positional uncertainties shown to lie well within
%the respective PSFs. PanchromaticPhotom has now been deprecated in light of
%the \textsc{lambdar} package \citep{wright16}, which produces contiguous photometry
%using r-band-defined apertures convolved with appropriate PSFs combined with a flux
%sharing algorithm for overlapping apertures.  
%It is included in this release for completeness and verification of earlier publications.

\subsubsection{LambdarPhotometry}

This DMU provides far-UV to far-IR aperture-matched photometry in 21-bands, measured consistently
using the Lambda Adaptive Multi-Band Deblending Algorithm in R (\textsc{lambdar}; \citealt{wright16}). 
Photometry has been measured using apertures defined as described in \citet{wright16}, 
where a considerable effort has been made to clean the input catalogue of apertures
affected by contamination or extraction problems such as shredding. Additionally, the photometry is
deblended from both GAMA sources and catalogued contaminants, which are defined independently for
the far-UV to near-IR data, the mid-IR data, and the far-IR data. Fluxes from this DMU show
considerable improvement in panchromatic consistency, i.e.\ smoothly changing behaviour 
with wavelength, when compared to the catalogue-matched photometry 
presented in \citet{driver16}.

\subsubsection{VisualMorphology}

This catalogue provides a number of visual morphological
classifications performed for various samples of
galaxies in the equatorial survey regions.
In total, 38\,795 sources have one or more classifications. 
\begin{itemize}
\item A basic visual classification using $giH$ images 
  (SDSS and UKIDSS/VIKING) was made by 2 team members for 
  sources with $0.002<z<0.1$. 
  The classes used were: Elliptical, NotElliptical, Little 
  Blue Spheroid (LBS), Star, Artefact and Uncertain. 
  A earlier version of the classification was used in \citet{driver12}. 
\item Hubble type galaxy classifications were made by 
  6 team members for sources with $0.002<z<0.06$. These were made using 
  a decision tree, which was translated to the following classes: 
  E, S0-Sa, SB0-SBa, Sab-Scd, SBab-SBcd, Sd-Irr, LBS, Star, Artefact. 
  See \S\,3 of \citet{kelvin14} and \S\,3.1 of \citet{moffett16} 
  for details. 
\item Disturbed galaxy classifications were made by 24 team members
  with multiple inspections of galaxies in close pairs and a control sample,
  over a redshift range of 0.01--0.33 depending on stellar mass. 
  Inspections were made of inverted $grK$ 
  images of 60\,kpc $\times$ 60\,kpc around each source using 
  SDSS and VIKING data. The classes used were: Disturbed, Normal, 
  and Unsure. See \S\,2.5 of \citet{robotham14} for details. 
\item A positional match was made to Galaxy Zoo~1 data 
  \citep{lintott11} for galaxies with $0.002<z<0.1$. 
  The columns included in this DMU are: 
  P\_EL, P\_CS, P\_EL\_DEBIASED, P\_CS\_DEBIASED. 
  The first two give the raw fraction of Galaxy Zoo votes for 
  Elliptical and NotElliptical, respectively, with the second 
  two giving the values corrected for redshift bias. 
  See \citet{lintott11} for details. 
\end{itemize}

\subsection{Spectral energy distribution DMUs}

These DMUs make use of photometric measurements and redshifts to derive
rest-frame luminosities, stellar population and dust properties. 

\subsubsection{kCorrections}

K-corrections are provided in the GALEX FUV and NUV bands, the
SDSS $ugriz$ bands and the UKIDSS $YJHK$ bands for all galaxies with
redshifts in the GAMA equatorial survey regions (i.e.\ excluding G02).
As well as the k-corrections themselves, we provide fourth-order polynomial fits
to K($z$), the k-correction as a function of redshift, to aid in the calculation of Vmax values.
These k-corrections are calculated using the eigentemplate fitting code \textsc{kcorrect} v4\_2 of 
\citet{BR07}, and are further described in \citet{loveday12}.

New for DR3, is that we provide k-corrections in passbands blueshifted by $z_0 = 0.2$ as well as 0.1 and 0.0.
\newtext{The advantage of using k-corrections for blueshifted passbands 
is that the uncertainties in the rest-frame magnitudes
are smaller when the choice of bandpass shift is similar 
to the typical redshifts of a galaxy sample 
(cf.\ Fig.~\ref{fig:hatlas-absmag-z}).}
We also provide k-corrections from fits to GAMA AUTO magnitudes (\S\,\ref{ApMatchedPhotom}), 
in addition to those from SDSS model magnitudes.
See \S\,2 of \citet{loveday15} for a discussion of the advantages of using 
GAMA AUTO magnitudes for this purpose.

\subsubsection{StellarMasses}

The StellarMasses DMU comprises estimates of total stellar mass and other population parameters
(including mean stellar age, dust attenuation, etc.) based on stellar population synthesis (SPS) of
the optical-to-near infrared SEDs. The
fits are based on the \citet{BC03} simple stellar population models, assuming a \citet{Chabrier03} IMF, 
uniform metallicity, exponentially-declining star formation histories, and single-screen
\citet{calzetti00} dust.  The parameter estimation is done in a Bayesian way (rather than, for example,
naive maximum likelihood), which has an important systematic effect on the inferred values for mass
and mass-to-light ratio (M/L).
The main reference for this DMU is \citet{taylor11}.  

The main improvement in the current version of this catalogue in comparison to DR2 is the
incorporation of the VISTA-VIKING near-IR photometry, which largely removes the problems seen in 
\citet{taylor11} when using the UKIDSS near-IR data. 
In the fitting, the full $ugrizZYJHK$ SEDs are weighted
in such a way that the fits are to a fixed restframe wavelength range of 3000–-11000\,\AA;
i.e.\ restframe $u$-—$Y$.  This decision is designed to protect against redshift-dependent biases arising
from, for example, the different availabilities of restframe near-UV information for galaxies at
different wavelengths.  In practice, there are not significant redshift-dependent systematic
differences between the current version of the StellarMasses DMU and earlier iterations.

\begin{figure}
\includegraphics[width=\singlecolsize\textwidth]{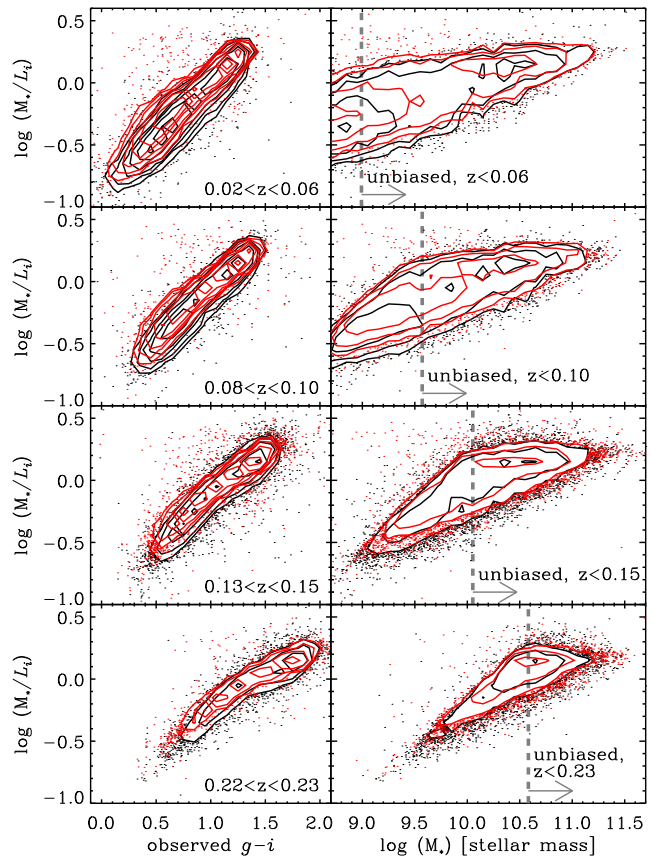}
\caption{\newtext{Mass-to-light ratios for four redshift slices versus 
  observed colour (left) and stellar mass (right). 
  Solar units are used for M/L and stellar mass. The red contours
  and points show stellar masses from the StellarMasses DMU 
  \citep{taylor11} while the black contours show the MagPhys DMU 
  \citep{driver17}. The vertical dashed lines represent the 
  lower limits in stellar mass for unbiased, 
  volume-limited, samples for the four different redshift limits. 
  Note the stellar 
  masses use \textsc{lambdar} photometry whereas the samples 
  are selected to $r_{\rm Petro} < 19.8\,\mathrm{mag}$.}}
\label{fig:ml-comparison}
\end{figure}

\newtext{\citet{taylor11} have shown that the numerical values of the SED-fit mass estimates can 
be approximated 
using the simple prescription (solar units):
   $\log (M_{*}/L_i) = -0.68 + 0.70 (g - i)_{\rm rest}$ 
with a typical 1-sigma uncertainty of $\sim0.12$\,dex.  This provides authors with a simple way of deriving robust mass 
estimates from minimal information.  Further, this provides a transparent way for authors to compare directly 
to the GAMA mass scale, in order to identify or account for possible systematic biases in the derivation of stellar mass estimates.
Figure~\ref{fig:ml-comparison} (left) shows M/L versus observed $g-i$. There is still a tight correlation becoming
less steep at higher redshift. Stellar mass derived using observed $g$ and $i$ magnitudes was used by 
\citet{bryant15} (their eq.~3) for a transparent stellar-mass selection
avoiding the need for even SED-fit k-corrections.}

\newtext{Figure~\ref{fig:ml-comparison} (right) shows M/L
in the $i$-band versus stellar mass for four redshift slices. 
This demonstrates the typical range in fitted M/L is from 0.2 to 2. 
At higher redshifts, GAMA samples becomes increasingly 
biased toward lower M/L because of the $r$-band magnitude selection limit.
In order to create volume- and stellar-mass limited samples, 
one needs to select the unbiased 
region in M/L as demonstrated by the vertical lines in the figure. 
These limits were obtained using a method similar to \citet{lange15}, 
see their fig.~1. Basically one needs to ensure that galaxies of 
every type (star-formation history, dust, profile), 
given a lower stellar-mass limit, 
could be selected over the redshift range of the sample.
In practice, it is not always necessary to be quite so strict 
and one could relax the stellar-mass limits shown by $\sim0.1$\,dex.}

\subsubsection{MagPhys}

The MagPhys DMU is based on parsing the extinction-corrected flux and flux error for colours from
\textsc{lambdar} through \textsc{magphys} \citep{dCCE08}.  The code fits stellar and dust-emission
templates to photometry, with dust attenuation applied to the stellar templates such that there is a
balance between the attenuated and the dust-emitted energy.  Energy balance is correct on average
for a random distribution of galaxy inclinations, but will overestimate (underestimate) attenuation
for face-on (edge-on) disks because of anisotropic attenuation.
The MagPhys DMU provides estimates of a number of key
parameters in particular the stellar mass, dust mass, and
star-formation rate (SFR), which we consider reliable and useful for broader
science.
The DMU is described in detail in \citet{driver17}.

\newtext{This DMU provides stellar mass estimates in addition 
to the StellarMasses DMU \citep{taylor11}. Both use 
\citet{BC03} models for the stellar population synthesis and assume 
a \citet{Chabrier03} IMF. For the MagPhys DMU, the dust 
attenuation uses the \citet{CF00} model with 
absorption redistributed in wavelength assuming various dust 
components for dense 
stellar birth clouds and for the ambient inter-stellar medium \citep{dCCE08}.
Fits are performed to the 21-band \textsc{lambdar} photometry using
models with a range of exponentially-declining star-formation histories, 
with bursts, and a range of dust attenuation. 
Note however that that the key flux at 250\,\micr\ 
is only measured with a S/N $> 2$ for about 30\% of galaxies. 
Figure~\ref{fig:ml-comparison} (left) shows a comparison of the M/L, 
for the different stellar mass estimates,
as a function of observed $g-i$. The median 
logarithmic offset 
varies from $-0.09$ for the lowest redshift slice, 
to $-0.03$ for the highest redshift slice shown.
The good agreement between the stellar mass estimates demonstrates
that the choice of dust attenuation approach does not make
a significant difference.}

\newtext{The long-wavelength baseline (UV to far-IR) used by the MagPhys DMU
also allow an estimate of SFR averaged over timescales less than a Gyr. 
MagPhys calculates SFRs from a combined UV and total IR SED fit,
summing both the unobscured and obscured star formation, and provides
various estimates over different timescales. For estimates of the SFR
over the last 0.1 Gyr, the formal 1-sigma errors typically range from
0.2\,dex to 0.6\,dex.\footnote{\newtext{The median estimate of 
the logarithm of the SFR over the last $10^8$\,yr is given by the 
column sfr18\_percentile50 for each galaxy in the MagPhys DMU. 
An estimate of the 1-sigma error was determined 
using (sfr18\_percentile84 $-$ sfr18\_percentile16)/2.
This encomposes measurement and fitting errors.}}
Detailed comparisons between these MagPhys SFRs
and other SFR indicators in GAMA are discussed in \citet{davies16,davies17},
and are used to determine the cosmic star-formation history in \citet{driver17}.}

\subsection{Environment DMUs}

\subsubsection{GroupFinding}

This DMU provides the GAMA Galaxy Group Catalogue (G3C) for the GAMA 
equatorial and G02 survey regions. The G3C is one of the major data products
for the GAMA project. At the most basic level it is a
friends-of-friends (FoF) group catalogue that has been run on GAMA
survey style mocks to test the quality of the grouping and then
run on the real GAMA data in order to extract our best effort groups. 
The details of the process are discussed extensively by \citet{robotham11}.

There have been a number of minor changes since the first version used by the GAMA team. 
Redshifts now use the CMB frame instead of the heliocentric frame. 
We include galaxies with lower-quality $nQ=2$ redshifts 
(\textsc{autoz} calibrated probability of 0.4 to 0.9) 
because if these galaxies link with a group, they are likely at the correct redshift 
since the chance of accidentally being aligned with a group is small. 
The grouping parameters were re-calibrated over a larger suite of GAMA mock lightcones \citep{farrow15} 
created using the \citet{gonzalez-perez14} model.\footnote{\newtext{The 
GAMA lightcone mocks are available through the Virgo database portal 
at virgodb.dur.ac.uk~(GAMA\_v1 table). 
They are built on the Millennium MR7 dark-matter-only simulation \citep{guo13} 
and are populated with galaxies following the 
\citet{gonzalez-perez14}, and an early version of the \citet{lacey16},
semi-analytic galaxy formation models.}}

The parameters are $b_0=0.0608$, $R_0=18.5000$, $E_b=0.0449$, $E_r=-0.0712$, and $\nu=0.7830$
as defined in \S\,2.1 of \citet{robotham11}. 
These determine the linking length in the radial and line-of-sight directions
as a function of survey location. 

The impact of these new grouping parameters is ultimately very small, with
grouping changes at the $\sim 1$\% level. A bigger update has been to use the \citet{farrow15}
randoms catalogue to determine the global redshift volume density [$n(z)$] rather than the
luminosity function fit originally used in \citet{robotham11}. 
Out to redshift $z=0.3$, the difference between the two
methods for estimating the local density of observable points is small, but at higher redshifts they smoothly
diverge at the 10\% level, resulting in larger implied densities and smaller implied
FoF links. The impact of this change is still fairly minor, but it does result in
fewer grouped high-redshift galaxies.

\subsubsection{FilamentFinding}

Elements of large-scale structure are clearly visible in GAMA
data. The FilamentFinding DMU identifies and characterizes filaments
and voids using a multiple-pass modified minimum spanning tree (MST)
algorithm. Initially, group centres from the GroupFinding DMU are used
as nodes for a MST that identifies filaments. All galaxies within a
distance $r$ of each filament are said to be filament galaxies
associated with that structure. A second MST is generated from all
galaxies that are not associated with filaments; this identifies
smaller-scale interstitial structures dubbed as `tendrils' \citep{alpaslan14}, 
which typically contain few galaxies and exist within
underdense regions. Galaxies that are beyond a distance $q$ from a
tendril are said to be isolated void galaxies. $r$ and $q$ are
selected so as to minimize the volume-weighted two point correlation
function of the void galaxy population. For further details, see
\citet{alpaslan14,alpaslan14B}. 

\subsubsection{GeometricEnvironments}

This DMU identifies the cosmic web of large scale structure within the
GAMA equatorial survey regions by classifying the geometric
environment of each point in space as either a void, a sheet, a
filament or a knot.
The classification system is based on evaluation of the deformation
tensor (i.e.\ the Hessian of the gravitational potential) on a grid. 
The number of eigenvalues
above an imposed threshold indicates the number of collapsed
dimensions of structure at that location -- either 0, 1, 2 or 3,
corresponding to a void, sheet, filament or knot, respectively. The
classification of the grid, as given in the GeometricGrid table,
allows the geometric environment of any object within the grid (any
object in the G09, G12 or G15 regions with $0.04 < z < 0.263$) to be
determined by assigning the object the same environment as the cell of
the grid in which the object is located.
See \citet{eardley15} for full details of the DMU. 

\subsubsection{EnvironmentMeasures}

The EnvironmentMeasures DMU provides three different metrics of the local
environment of GAMA galaxies. The three different metrics are the 5th
nearest-neighbour surface density \citep{brough13}, the number of galaxies
within a cylinder \citep{liske15}, and the adaptive Gaussian environment parameter \citep{yoon08}. 
The method used to calculate these has not changed from DR2 to DR3.

\subsubsection{Randoms}

The random catalogue DMU contains a series of randomly-placed points
(`randoms'), each tagged with a galaxy CATAID, that fill the
volume of the equatorial fields of the GAMA survey in a way that
follows the galaxy selection-function. It is designed to be used as a
reference for measuring overdensities and estimating clustering
statistics. See \citet{farrow15} for details.

The method used to produce these points was introduced in \citet{cole11}, whilst the particular
implementation for GAMA is given in \citet{farrow15}. The method involves cloning real galaxies,
with each random point in the catalogue being stored with the CATAID of its parent. This
allows the randoms to be assigned galaxy properties by matching on CATAID. Once properties
are assigned to the randoms a suitable random catalogue can be produced by applying the same sample
selection cuts as applied to the galaxies. One must, however, take care when using
luminosity-selected samples, as a particular form of luminosity evolution is explicitly assumed in
their production. In addition, if your sample has additional
sources of incompleteness beyond the $r <19.8$ selection, for example a H${\alpha}$ flux limit,
further work is required so please contact the DMU authors 
(for an example of this, see Gunawardhana et al.\ in preparation).

The DMU contains two tables. In {\sc RandomsUnwindowed} the standard \citet{cole11} method is used,
whilst in the {\sc Randoms} table the randoms created from a cloned galaxy are restricted to a
window around the redshift of that galaxy, as explained in \citet{farrow15}. The window is
designed to minimise the impact of any unmodeled galaxy evolution effects on the random $n(z)$.

The DMU version of the randoms has two important differences to the ones used in \citet{farrow15}. 
Firstly, the parameters used to model luminosity evolution have been changed following the
results of \citet{loveday15}. Secondly the size of the redshift window has been increased, to
adjust the balance between limiting the effect of unmodeled galaxy-evolution and keeping the window
large enough to completely remove large-scale structure from the input catalogue.

\section{Data Release 3}
\label{sec:summary}

The third GAMA data release (DR3) provides AAT/AAOmega spectra, redshifts and a wealth of
ancillary information for \newtext{primarily} 215\,260 objects from the GAMA~II survey (Table~\ref{tab:gamaregions}).
Of these, 178\,856 are main survey objects and 36\,404 are fillers. 
In turn, 150\,465 (84\%) and 4\,344 (12\%) of these, respectively, have secure redshifts. 
DR3 updates all data previously released in DR2, and significantly expands on DR2: 
DR3 includes both more objects and a wider range of data products than DR2. 
DR3 thus supersedes DR2 in every way. 
The data release is available at http://www.gama-survey.org/dr3/~.

In DR3 we are releasing data \newtext{primarily} 
for the following GAMA~II objects:
\begin{itemize}
  \item Main survey objects: all for G02 and G15, $r < 19.0\,\mathrm{mag}$ 
selected for G09 and G12;
  \item Fillers: all for G02, H-ATLAS selected for G09, G12 and G15.
\end{itemize}
The region areas and main survey limits for GAMA~I, GAMA~II and DR3 are shown in
Table~\ref{tab:gamaregions}.  In addition, we include a small set of objects that were part of DR2
or the H-ATLAS DR1 and not already covered by the selection above. Note that for G02, we are
releasing all GAMA~II data, and for G15, all main survey data. The environment
measurements that use the equatorial main survey are only available for G15.

\begin{figure}
\centerline{
\includegraphics[width=\singlecolsize\textwidth]{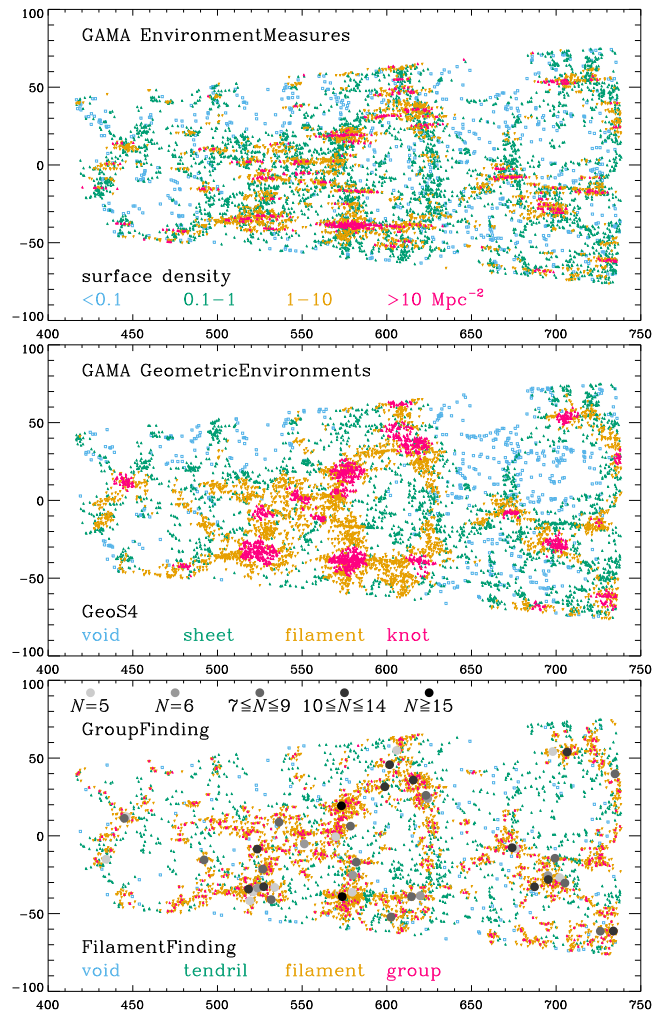}
}
\caption{Galaxy distribution in the GAMA G15 field showing
  different environmental measurements. Galaxies were selected
  with $0.10<z<0.18$ and $-1.5\degr < \mathrm{dec} < 1.5\degr$
  (the two dimensions shown correspond to RA and redshift). 
  Each panel shows the galaxies colour coded according to environment: 
  top panel, using the 5th nearest neighbour surface density \citep{brough13}; 
  middle panel, using the geometric environment classifications \citep{eardley15}; 
  lower panel, using the filament-finder classifications \citep{alpaslan14}.
  The top two panels show a volume-limited sample with 
  $M_r<-20.0\,\mathrm{mag}$, 
  while the filament finder uses only galaxies with 
  $M_r < -20.55\,\mathrm{mag}$ ($h=0.7$). 
  In the latter case, the classified groups, from the FoF catalogue \citep{robotham11}, 
  used as the starting point of the filament finder, 
  are those with two or more members from the volume-limited sample. 
  Also shown are rich groups shaded according to the number of member 
  galaxies with $M_r < -21.0\,\mathrm{mag}$
  (cf.\ Fig.~\ref{fig:cone-g02}). 
  Note also the FilamentFinding DMU provides links beween galaxies and groups in a filament.}
\label{fig:enviros}
\end{figure}

GAMA DR3 doubles the number of galaxies with released spectroscopic redshifts
compared to DR2 \citep{liske15}. New redshifts are available, in particular,
for the G02 region (\S\,\ref{sec:target-select}), H-ATLAS selected sources
(\S\,\ref{sec:h-atlas}), and in G15. The redshift measurements now
use the more accurate \textsc{autoz} code \citep{baldry14}. 
New environment measurements for G15 are made available including 
a filament catalogue \citep{alpaslan14}, 
and geometric measurements \citep{eardley15}. 
The galaxy distribution is shown in Fig.~\ref{fig:enviros} 
with colour coding to showcase the different environment 
\newtext{classifications}. 
New image analysis is made available including 
\textsc{lambdar} photometry \citep{wright16} that provides consistent
photometry across all the bands from the far-UV to far-IR. 
All the available redshifts, including the G23 region, 
and all GAMA data products will be made available in DR4. 

\section*{Acknowledgements}

GAMA is a joint European-Australasian project based around a
spectroscopic campaign using the Anglo-Australian Telescope. The GAMA
input catalogue is based on data taken from the Sloan Digital Sky
Survey and the UKIRT Infrared Deep Sky Survey. Complementary imaging
of the GAMA regions is being obtained by a number of independent
survey programmes including GALEX MIS, VST KiDS, VISTA VIKING, WISE,
Herschel-ATLAS, GMRT and ASKAP providing UV to radio coverage. GAMA is
funded by the STFC (UK), the ARC (Australia), the AAO, and the
participating institutions. The GAMA website is
http://www.gama-survey.org/ .

I.~Baldry acknowledges funding from Science and
Technology Facilities Council (STFC) and Higher Education Funding
Council for England (HEFCE).
L.~Dunne acknowledges support from the European Research Council advanced 
grant COSMICISM and also consolidator grant CosmicDust.
H.~Hildebrandt is supported by an Emmy Noether grant (No.\ Hi 1495/2-1) of the 
Deutsche Forschungsgemeinschaft.
P.~Norberg acknowledges the support of the Royal Society through the award
of a University Research Fellowship, and of the Science and Technology
Facilities Council (ST/L00075X/1). 

\setlength{\bibhang}{2.0em}
\setlength\labelwidth{0.0em}

%\bibliographystyle{mn2e-williams}
%\bibliography{galaxies,surveys,stars,general}

%%\bsp

\label{lastpage}

\end{document}